# Mechanical properties of viral capsids

Roya Zandi[1] and David Reguera[2]

[1]*Department of Chemistry and Biochemistry, University of California at Los Angeles, Los Angeles, California 90095-1569, USA*
[2]*Department de Física Fonamental, Facultat de Física, Universitat de Barcelona, Martí i Franquès 1, Barcelona 08028, Spain*

Viruses are known to tolerate wide ranges of *p*H and salt conditions and to withstand internal pressures as high as 100 atmospheres. In this paper we investigate the mechanical properties of viral capsids, calling explicit attention to the inhomogeneity of the shells that is inherent to their discrete and polyhedral nature. We calculate the distribution of stress in these capsids and analyze their response to isotropic internal pressure (arising, for instance, from genome confinement and/or osmotic activity). We compare our results with appropriate generalizations of classical (i.e., continuum) elasticity theory. We also examine competing mechanisms for viral shell failure, e.g., in-plane crack formation vs radial bursting. The biological consequences of the special stabilities and stress distributions of viral capsids are also discussed.

## I. INTRODUCTION

Viruses are amongst the simplest biological systems. They are essentially composed of a protein shell (or "capsid") that encloses the viral genetic material (RNA or DNA) [1]. Viral capsids consist of several copies of either one type of protein ("subunit") or of a few slightly different proteins. It is remarkable that many plant and animal viruses can be self-assembled *in vitro* from their purified RNA or DNA, and the isolated capsid proteins. At the right ionic strength and *p*H, the capsid proteins can form (empty) shells even in the absence of their genomes. One of the main purposes of a capsid is to protect its genome against adverse conditions existing inside a cell or in-between cells. To this end, the mechanical response of viral capsids to diverse environmental hazards, their stability, and the threshold of their mechanical failure and disassembly are major issues in ensuring the viability of viruses.

Recent micromechanical experiments illustrate the amazing mechanical properties of viral capsids. For instance, bacteriophage capsids have been shown to be dramatically strong, withstanding internal osmotic pressures as high as 100 atmospheres [2–6]. The cowpea chlorotic mottle virus (CCMV) is able to swell and maintain its structural integrity under wide ranges of *p*H and salt concentrations [7]. More recently, atomic force microscopy measurements revealed that the capsid shell of CCMV could recover from deformations as large as 30% before rupturing [8].

These rather unusual properties make viral capsids ideal candidates for a range of novel applications such as nanocontainers for drug delivery. As a consequence, the synthesis of viral nanocontainers has become a rapidly developing area of material science and soft condensed matter physics. Furthermore, the effect of hydrostatic pressure on the structure of viral particles has been investigated extensively due to its potential application in biotechnology and antiviral vaccines [9]. Thus the study of the mechanical properties of viral capsids is vital in understanding both the biological function of viruses and their potential performance as bionanomaterials. The aim of this paper is precisely to investigate the unique mechanical properties of viral capsids. We will examine their response to different kinds of external and internal influences, and evaluate their strengths and mechanisms of failure.

Despite the varieties in size, sequence, and conformation of capsid proteins, most spherical viruses self-assemble into capsids with icosahedral symmetry [10]. Figures 1(a) and 1(b) illustrate two examples: CCMV [11] (mentioned earlier) and Herpes simplex virus [12], respectively. The shell of CCMV consists of several copies of a single capsid protein while the shell of Herpes virus is composed of many copies of four different capsid proteins. Regardless of the diversity in the number of protein subunits, the proteins in the shells of these viruses, like in many other spherical viruses, are organized in two different structural units: aggregates of five ("pentamers") and six ("hexamers") proteins, known as "capsomers." As depicted in the picture, the capsomers are arranged in a two-dimensional closed shell in both cases. CCMV is made up from 12 pentamers and 20 hexamers while Herpes contains 12 pentamers and 150 hexamers.

The classical study of Caspar and Klug (CK) [13] introduced the idea of quasiequivalence to justify the special architectures adopted by viruses, and laid the foundation of modern structural virology. They showed that closed shells with icosahedral symmetry can be constructed from a hexagonal lattice by inserting pentagons at sites of the hexagonal lattice that are separated by $h$ steps along one lattice direction and $k$ along the other, with $h$ and $k$ non-negative integers. The integer $T=h^2+k^2+hk$ is then the number of inequivalent sites in the resulting shell, which is made of 12 pentagons and $10(T-1)$ hexagons. The total number of protein subunits is thus $12\times 5 + 10\times(T-1)\times 6 = 60T$. The CK classification has been shown to apply to a huge number of spherelike viruses. Figure 2 shows two examples of CK structures. The CCMV and Herpes pictures in Fig. 1 have the same symmetry and number of pentamers and hexamers as the $T=3$ and $T=16$ structures in Fig. 2, respectively.

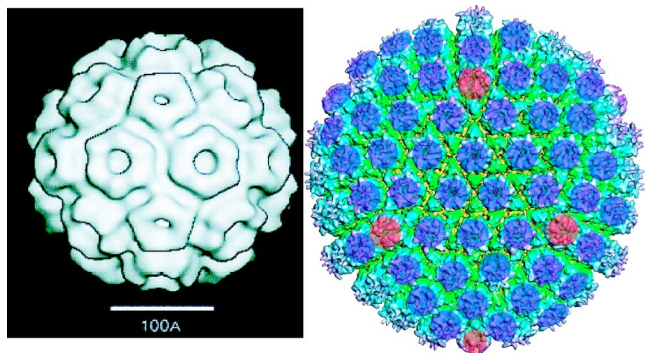

FIG. 1. (Color online) Cryo-TEM reconstruction of (a) CCMV and (b) Herpes simplex virus. The figures are not to scale; the diameter of Herpes virus is almost 5 times bigger than that of CCMV.

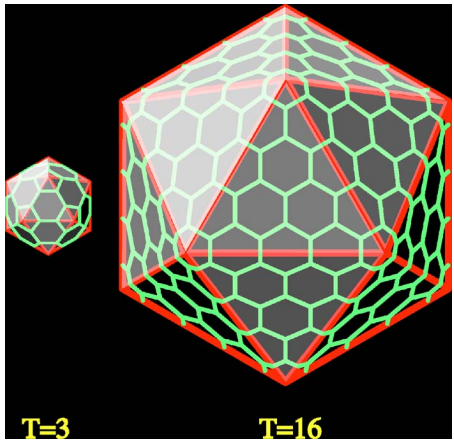

FIG. 2. (Color online) CK organization for (a) $T=3$ and (b) $T=16$ structures.

Concepts from continuum elasticity theory have already been used to characterize some features of the stress distribution in viral capsids. According to continuum elasticity theory, icosahedral shells are characterized by a stress pattern focused on 12 vertices of fivefold sites, which can be considered as disclinations in a 2D hexagonal network. To construct in-plane strain-free structures with icosahedral shape from a hexagonal sheet (Fig. 2), one has to impose curvature on the shell along the edges connecting the fivefold sites. This pronounced curvature seems to be absent among hexamers connecting two pentamers in CCMV and herpes [Figs. 1(a) and 1(b)], which are quasispherical. This implies the presence of a significant in-plane elastic stress in the CCMV and Herpes shells. Indeed, there are two competing elastic energies in a spherical shell: strain and bending energies. The competition between these two energies suggests a dimensionless quantity that is useful in understanding the shape of spherical shells: the Foppl–von Karman number, $\gamma = YR^2/\kappa$ [14]. Here $R$ is the radius of the sphere, $Y$ is the in-plane (2D) Young's modulus, and $\kappa$ is the bending rigidity of the spherical shell. If the value of Foppl–von Karman number exceeds 154, a buckling transition takes place and sharp corners appear connecting fivefold sites. Thus in the case of viruses, as the size of the shell increases, the continuum theory predicts that eventually a buckling transition takes place and that the spherical capsid facets into the polyhedral shapes similar to the ones shown in Fig. 2 [14]. Note that if the bending rigidity $\kappa$ between capsomers is very big, or the Young's modulus very small, this transition will only take place for very large spheres—the critical radius for buckling grows as $R \sim \sqrt{\kappa/Y}$. The focus of this work is on understanding the in-plane stresses of spherically shaped viruses and not on the faceted shaped viruses.

The existence of two different structural units—pentamers and hexamers—which can also adopt different positions in the capsid, already indicates that the mechanical properties, and in particular the stress distribution of viral shells, are inhomogeneous. We will see later that the stress distribution is conditioned by the particular geometry of the shell and by the fact that it is constituted by a discrete number of proteins organized into a one-protein-thick shell. It is thus not obvious that all the concepts that are valid for macroscopic materials and that are introduced by continuum elasticity theory are readily applicable to small viral shells consisting of a limited number of proteins.

In fact, the difficulties in the extrapolation of continuum arguments to describe the stress at the molecular scale have already been noticed in the analysis of systems geometrically very similar to viruses [15]. Fullerenes and carbon nanotubes share with viruses a similar geometrical structure, e.g., $C_{60}$ has the same architecture as a $T=3$ capsid, and carbon nanotubes resemble the structure of rodlike viruses. The mechanical properties of fullerenes and carbon nanotubes have been recently investigated from simulation, theory and experiments [16]. Some of the strategies developed to study these systems could be potentially useful in the analysis of the mechanical properties of viral capsids; however, one has to consider that the carbon atoms in fullerenes and carbon nanotubes are held together by strong, directional, covalent bounds rather than by the noncovalent interactions of viral proteins.

The issue of elastic stress is fundamental to the study of stability of a structure. Perfectly homogeneous materials fail when the elastic stress exceeds a threshold value. *Heterogeneous* materials fail at much lower stresses, however, due to the focusing of stress near local heterogeneities, resulting in the nucleation of cracks. It would be reasonable to infer that the stability of viral capsids could be determined by the condition that local stresses have to remain below a fracture threshold.

The aim of this paper is to examine the mechanical properties of viral shells and to learn if an understanding of the elastic stress of viral capsids gives us a better insight into their biological functions. The rest of the paper is organized as follows. In Sec. II we present our model, which captures successfully the essential elements of a capsid structure. Using this model, we compute, in Sec. III, different quantities characterizing mechanical properties of viral shells. Section IV is devoted to the analysis of the distinct features that arise from the unique nature of these nanoshells, in particular from the discrete character and from the intrinsic inhomogeneity imposed by the geometry and curvature of viral shells. In Sec. V, we analyze the effect of pressure (imposed externally or internally, for instance, by a confined material or osmolytes) on the shell. We investigate, in Sec. VI, how this pressure is related to lateral failure or bursting of capsids. Finally, in Sec. VII, we summarize the main results of our work and discuss the relevance of mechanical stability of viral capsids to the different plausible genome release scenarios.

## II. MODEL OF CAPSID SELF-ASSEMBLY

To investigate the equilibrium structure of viral capsids, we have recently presented a model that has been able to reproduce successfully, as free energy minima, the $T$-number structures as well as other (nonicosahedral) structures observed *in vitro*. Irrespective of the sequence and structural differences of capsid proteins, for a wide range of viruses such as CCMV, Polyoma, HIV, and HK97, capsomers appear as common and well-defined structural units. To this end, in our model we do not attempt to describe individual protein subunits (whose interactions are expected to be asymmetric and species-specific) but rather we focus on the *capsomers*, which interact with a more isotropic and generic interaction.

Our model captures the essential ingredients of capsomer-capsomer interaction in the self-assembly process: a short-range repulsion which describes the rigidity of the capsomers, i.e., resisting compression, and a longer-range attraction representing hydrophobic attractions between capsomers. We assume that the effective capsomer-capsomer potential $V_{ij}(r)$ depends only on the separation $r$ between capsomer centers. Specifically, we use the following Lennard-Jones potential

$$V_{ij}(r) = \varepsilon_0\left(\left(\frac{\sigma_{ij}}{r}\right)^{12} - 2\left(\frac{\sigma_{ij}}{r}\right)^6\right), \quad (1)$$

which satisfies the above-mentioned requirements. In Eq. (1) $\varepsilon_0$ is the capsomer-capsomer binding energy, which is taken to be $15k_BT$, a typical value reported from atomistic calculations of subunit binding energies [18] (here $k_B$ is the Boltzmann constant and $T$ the absolute temperature). Capsomers are constrained to be on the surface of a sphere (the "capsid") of radius $R$. The quantity $\sigma_{ij}$ is the distance between the centers of capsomers at edge contact (which obviously depends on the radius of the capsid).

As illustrated in Fig. 1, an essential feature of viral capsids is the existence of two different morphological units (pentamers and hexamers). To account for this we assume that capsomers can adopt one of two internal states: $P$ (entamer) and $H$ (examer). The potential has the same form for interactions between different capsomer types except that the equilibrium spacing $\sigma_{ij}$ includes the geometrical size difference between pentamers and hexamers. Explicitly, the ratio of sizes of pentamers vs hexamers was chosen as $\sin 30°/\sin 36° = 0.85$, which corresponds to the ratio of circumscribed radii of a hexagon and a pentagon with the same edge length. We also introduce an energy difference $\Delta E$ between a $P$ and an $H$ capsomer, which reflects specific differences that may exist between protein contact interactions and folding conformations of pentamer and hexamer proteins.

We have carried out Monte Carlo simulations in which $N$ interacting capsomers are allowed to range over a spherical surface of radius $R$. For each value of $N$ and a fixed radius $R$, we have evaluated the equilibrium configurations and their corresponding average energy. We have then repeated the procedure for different values of $R$, to get the optimal value of the radius $R^*$ that minimized the energy of $N$ capsomers. For each fixed total number of capsomers $N$, the number $N_P$ of $P$ units (and hence the number $N_H = N - N_P$ of $H$ units) are permitted to vary and is not fixed to be twelve (as in the CK construction). More explicitly, during equilibration, we have allowed switching between $P$ and $H$ states, with a Boltzmann probability factor $\exp^{-\beta\Delta E}$, thus exploring all possible *geometries* and *conformations*. As the number of capsomers increases, it becomes harder and harder to obtain the optimal structures with a plain Metropolis scheme. The number of local minima increases and the system can get easily stuck in metastable configurations. To overcome this difficulty, we have used simulated annealing [19] to obtain the putatively optimal structures of big capsids [20].

Our simulations indicate that for special magic numbers of capsomers the energy per capsomer exhibits a deep minimum. If the $P$ to $H$ switching energy $\Delta E$ is small compared to the capsomer binding energy $\varepsilon_0$, then the magic numbers correspond to the Caspar-Klug $T$-number structures: $N = 12\ (T=1), 32\ (T=3), 42\ (T=4), 72\ (T=7)$, etc. The number of pentagons in these optimal structures is always exactly 12, and each configuration has icosahedral symmetry, with 12 fivefold sites (pentamers) and $10(T-1)$ sixfold sites (hexamers). Consequently, we have shown that the $T$-number structures arise naturally as free energy minima of our model [17].

It is important to note that there are two further ingredients that play important roles in self-assembly process. One is the chemical potential which controls when the self-assembly of the capsid is energetically favorable compared to remaining as free proteins or capsomers in solution. The other element is the spontaneous curvature and the related bending energy of the capsomer-capsomer interaction, whose main role is in the selection of one structure (one of the free energy minima) from the various optimal $T$-number structures. We will elaborate later on the relevance of these two factors on the mechanical behavior and disassembly of the capsids.

In the following section, we use thermally equilibrated $T$-number structures, obtained by Monte Carlo simulations, as a basis for an analysis of the stress pattern and the mechanical properties of capsids. Note that in addition to the Lennard-Jones interaction discussed above, we have tested different forms for $V(r)$ and have found that structures with icosahedral symmetry arise as long as we use potentials with a range of attraction comparable or longer than the one given in Eq. (1). In this context, a more sophisticated intercapsomer potential has been recently introduced and shown to describe properly the equilibrium structure of a $T=1$ virus [21]. It is obvious that the *actual* values of stresses in the equilibrated structures strongly depend on the strength and details of the interaction. In this respect, it is not our intention to obtain *quantitative* agreement with real capsids. However, the *qualitative* results obtained here are expected to be general to all interactions sharing the same characteristics (i.e., short-range repulsion and a longer-range attraction).

## III. STRESS TENSOR OF $T$-NUMBER CAPSIDS

### A. Global definition of stress and pressure

The stress tensor is one of the most important quantities that characterizes the mechanical properties of a system. The *global* stress is commonly measured at the atomic level as the equilibrium average of $\sigma_{\alpha\beta}$, the virial stress tensor [22]

$$\Omega^d \sigma^d_{\alpha\beta} = \sum_i \left( \sum_j \frac{1}{2} m_i v_i^\alpha v_j^\beta - \frac{1}{2} \sum_{j \neq i} \frac{dV_{ij}(r_{ij})}{dr_{ij}} \frac{r_{ij}^\alpha r_{ij}^\beta}{r_{ij}} \right). \quad (2)$$

Here, $m_i$ and $v_i^\alpha$ denote the mass and the $\alpha$ component of the velocity of particle $i$, $V_{ij}(r_{ij})$ is the interaction potential, $\alpha, \beta$ denote the components in an orthogonal system of coordinates, $i$ and $j$ are the particle indices, and the summation is over all particles in the system occupying a volume $\Omega^d$ ($\Omega^d$ represents the area in $d=2$ dimensions, and the volume in $d=3$ dimensions). The right-hand side of the above equation accounts for two different contributions to the stress. The second term is essentially the sum of the product of the force in one of the orthogonal directions times the separation between each pair of capsomers along another (or the same) direction. The first term is a kinetic contribution originating from the movement of the particles. For the present case, where the thermal energy is small compared to the characteristic interaction energy, the first term can be neglected when the system is at thermal equilibrium.

We will use a spherical coordinate system appropriate for a system confined to the surface of a sphere. Because of the quasi-two-dimensional (2D) nature of our one-protein-thick viral shells, the 2D stress tensor $\sigma^{2D}_{\alpha\beta}$ will be used to discuss the lateral strength of the capsid. Specifically, we will decompose the stress tensor into three parts and compute the lateral stress

$$\sigma^{2D}_T = -\frac{1}{4\pi R^2} \frac{1}{2} \sum_{i,j} \frac{dV(r_{ij})/dr_{ij}}{r_{ij}} \frac{(r_{ij} \cdot \hat{e}_\theta)^2 + (r_{ij} \cdot \hat{e}_\phi)^2}{2}, \quad (3)$$

the (45°) shear stress

$$\sigma^{2D}_{\theta\phi} = -\frac{1}{4\pi R^2} \frac{1}{2} \sum_{i,j} \frac{dV(r_{ij})/dr_{ij}}{r_{ij}} (r_{ij} \cdot \hat{e}_\theta)(r_{ij} \cdot \hat{e}_\phi), \quad (4)$$

and the elongational shear stress

$$\sigma^{2D}_E = -\frac{1}{4\pi R^2} \frac{1}{2} \sum_{i,j} \frac{dV(r_{ij})/dr_{ij}}{r_{ij}} \frac{(r_{ij} \cdot \hat{e}_\theta)^2 - (r_{ij} \cdot \hat{e}_\phi)^2}{2}. \quad (5)$$

Here, $\hat{e}_\theta$ and $\hat{e}_\phi$ are the unit vectors along the polar and azimuthal directions, respectively, on the spherical surface. Both types of shear would break icosahedral symmetry, so globally $\sigma_E = \sigma_{\theta\phi} = 0$ for all CK structures.

It is important to note that the force on a capsomer, due to the interaction of neighboring capsomers, also has an out-of-plane, radial component that is not included in the two-dimensional stress tensor. This radial out-ward force would produce an effective reaction pressure on the spherical surface equal to

$$\Delta p = -\frac{1}{4\pi R^2} \sum_{i<j} \frac{dV_{ij}}{dr_{ij}} \frac{r_{ij} \cdot \hat{e}_r}{r_{ij}}. \quad (6)$$

The quantity $\Delta p$ is a measure of the *pressure difference* between the interior and the exterior of the capsid required to keep capsomers confined to the spherical surface; $\hat{e}_r$ is the unit tangent vector in the radial direction. The pressure difference $\Delta p$ translates into a distribution of lateral stresses and determines the lateral strength of the capsid; however, there is no obvious relationship between the mechanical definition of pressure given in Eq. (6) and that of the lateral stress in Eq. (3). To investigate the coupling between these two quantities, we resort to the thermodynamic definition of pressure:

$$\Delta p = -\left(\frac{\partial F}{\partial V}\right)_{N,T} = kT\left(\frac{\partial \ln Q}{\partial V}\right)_{N,T}, \quad (7)$$

where $F$ is the (Helmholtz) free energy of the system and $Q$ is its partition function. In the Appendix, we show that this definition leads to the following relationship between the pressure difference and the 3D global virial stress tensor from Eq. (2):

$$\Delta p = \frac{1}{3}\text{Tr }\sigma^{3D}. \quad (8)$$

Here Tr stands for the trace, and $\sigma^{3D}$ is the 3D stress tensor, whose lateral and radial components are explicitly given by

$$\sigma^{3D}_T = -\frac{3}{4\pi R^3} \frac{1}{2} \sum_{i,j} \frac{dV(r_{ij})/dr_{ij}}{r_{ij}} \frac{(r_{ij} \cdot \hat{e}_\theta)^2 + (r_{ij} \cdot \hat{e}_\phi)^2}{2}, \quad (9)$$

$$\sigma^{3D}_R = -\frac{3}{4\pi R^3} \frac{1}{2} \sum_{i,j} \frac{dV(r_{ij})/dr_{ij}}{r_{ij}} (r_{ij} \cdot \hat{e}_r)^2. \quad (10)$$

It is important to note that we have verified that both the mechanical and the thermodynamic definitions, namely Eqs. (6) and (8), provide the same value of the pressure for all the cases considered in this paper.

Note that, from the previous definitions of stress, the 2D and the 3D radial stresses in our spherical shell are related by

$$\sigma^{3D}_{R,T} = \frac{3}{R} \sigma^{2D}_{R,T}. \quad (11)$$

Thus one can rewrite the pressure in terms of the 2D lateral stress as

$$\Delta p = \frac{1}{3}(2\sigma^{3D}_T + \sigma^{3D}_R) = \frac{2\sigma^{2D}_T}{R} + \frac{1}{3}\sigma^{3D}_R, \quad (12)$$

which constitutes a kind of "generalized Laplace equation" accounting for the presence of additional radial stresses in our curved 2D system. Notice that, when the radial component of the stress becomes negligible, Eq. (12) reduces to the standard Laplace relation [23],

$$\Delta p = \frac{2\sigma^{2D}_T}{R}, \quad (13)$$

where the lateral 2D stress $\sigma^{2D}_T$ plays the role of a surface tension.

### B. Local definition of stress and pressure

As noted in the Introduction, fracture and failure of capsids are determined by *local* values of stress rather than average quantities since they are in general defined in homo-geneous systems. Thus in addition to global measures, a local definition of stress is also required. However, the definition of a local stress tensor for a microscopic discrete system is not trivial. The main problem lies in the macroscopic and continuum nature of the definitions of stress and strain, which become ambiguous at the molecular level. In fact, several definitions for local stresses have been proposed [15].

The most obvious prescription to calculate an atomic-level or local stress tensor $(\sigma^d_{\alpha\beta})_i$ is to assume that the previous bulk measure of stress, Eq. (2), would be valid for a small volume $\Omega^d_i$ around a particle $i$, that is

$$\Omega^d_i (\sigma^d_{\alpha\beta})_i = \sum_j \frac{1}{2} m_i v_i^\alpha v_j^\beta - \frac{1}{2} \sum_{j\neq i} \frac{dV_{ij}(r_{ij})}{dr_{ij}} \frac{r_{ij}^\alpha r_{ij}^\beta}{r_{ij}}. \quad (14)$$

The simplest choice of $\Omega^d_i$ is the volume per particle $\Omega^d_i = \Omega^d_{tot}/N$ (in our case, the area or volume per capsomer). However, this selection is only valid for a homogeneous system. In the presence of inhomogeneities, other prescriptions for defining the "atomic" volume for the stress become more reasonable. In fact, many alternative definitions of local stress have been proposed, including among others the atomic stress [24] and the Lutkso stress [25]. The actual values of local stresses might change using these different definitions. Nevertheless, since the focus of this paper is only on the qualitative behavior of stress tensors, we are not concerned about these differences. Thus we will use the simplest and most convenient definition of local stress, which is the local virial stress. More precisely, the local lateral or tangential 2D stress, which is the most relevant quantity to discuss the lateral strength and failure of capsids, is defined as

$$(\sigma^{2D}_T)_i = -\frac{N}{4\pi R^2} \frac{1}{2} \sum_j \frac{dV(r_{ij})/dr_{ij}}{r_{ij}} \frac{(r_{ij}\cdot\hat{e}_\theta)^2+(r_{ij}\cdot\hat{e}_\phi)^2}{2}. \quad (15)$$

We will also analyze the potential presence of (45°) shear stresses, defined locally as

$$(\sigma^{2D}_{\theta\phi})_i = -\frac{N}{4\pi R^2} \frac{1}{2} \sum_j \frac{dV(r_{ij})/dr_{ij}}{r_{ij}} (r_{ij}\cdot\hat{e}_\theta)(r_{ij}\cdot\hat{e}_\phi) \quad (16)$$

and evaluate the local radial component of the 3D stress tensor

$$(\sigma^{3D}_R)_i = -\frac{3N}{4\pi R^3} \frac{1}{2} \sum_j \frac{dV(r_{ij})/dr_{ij}}{r_{ij}} (r_{ij}\cdot\hat{e}_r)^2, \quad (17)$$

which is required to relate the pressure with the lateral stress through Eq. (12).

Armed with all these definitions, we are now ready to characterize the mechanical properties of the optimal structures resulting from our simulations.

## IV. MECHANICAL PROPERTIES OF $T$-NUMBER CAPSIDS

### A. Distribution of lateral stress

Figure 3 illustrates the distribution of local lateral stresses—see Eq. (15)—of some of the $T$-number optimal structures. The color scale on the plot specifies the values of stresses. There are several important points that can be inferred from the figure. We enumerate them as follows.

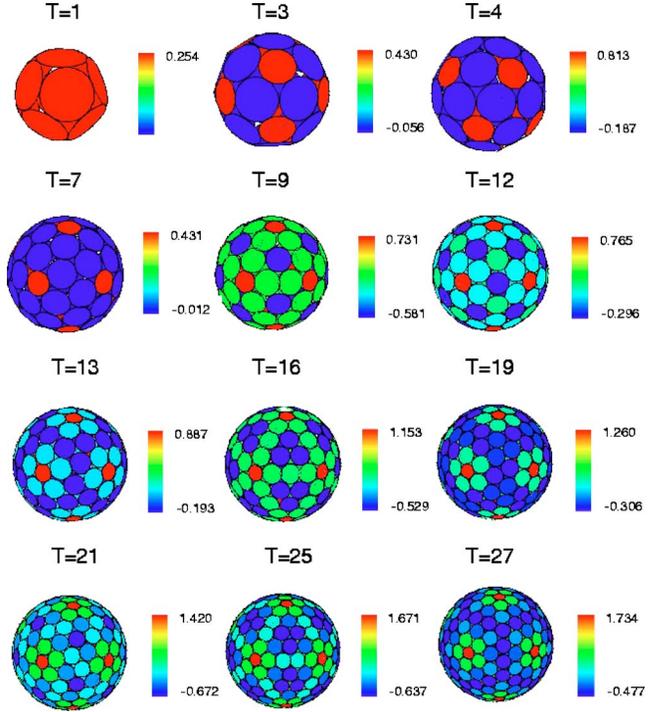

FIG. 3. (Color online) Lateral stress distribution for some of the $T$-number structures. Different colors indicate different levels of stress.

(1) Except for $T=1$, in which the stress distribution is obviously homogeneous, we observe that the stress distribution is inhomogeneous and the heterogeneity increases (although not monotonically) with $T$ numbers. For instance, for $N=32/T=3$, there are only two different stress levels: one corresponding to 12 (red) pentameric capsomers, and the other one to 20 hexameric capsomers (blue). On the other hand, for the largest icosahedral shell shown in the picture, $N=272/T=27$, there are five distinctly different stress levels.

(2) Pentamers and hexamers have different values of stress. Pentamers have consistently more (positive) stress than hexamers. In fact, Pentamers are under significant lateral compression in the optimal $(R=R^*)$ structures. The compressive stress on the pentamers increases with the size of the capsid, as predicted by continuum elasticity theory. The existence of compressional stress in the fivefold sites of the capsid seems to be a general "requirement" for icosahedral symmetry. In our simulations, we observed consistently that structures with icosahedral symmetry could be obtained from non icosahedral ones by compression $(R<R^*)$, for all structures in which the number of capsomers is $N=10\times T+2$. For instance, extremely short-ranged potentials (such as the "contact" potential used in Ref. [26] combined with single size capsomers (i.e., $|\Delta E|\gg 1$) may lead to optimal structures lacking icosahedral symmetry. However, if we impose a compressive strain on a structure (obtained by selecting a radius $R$ smaller than the optimal radius $R^*$), the equilibrated structure adopts icosahedral symmetry. The concentration of stress in the fivefold sites (the pentamers) seems to alleviate optimally the global stress of a structure by allowing hexamers to become essentially stress free. The compression of the capsomers at fivefold sites is in fact the driving force for the switch of a hexamer to a pentamer in our simulations. The smaller size of the pentamers alleviates the energetic cost of compression and reduces the total energy even further. It is worth noticing the similarities of this phenomenon with the absence of 12 carbon atoms at the icosahedral vertices in larger fullerenes, which has also been proposed to relieve strain [27].

(3) As depicted in the figure, there exist different stress levels in hexamers of the higher $T$-number structures. For the structures with smaller $T$ numbers, i.e., between $T=3$ and $T=19$, the distribution of lateral stress depends markedly on the geometry of the capsids, and there seems to be no general trend in how the stress is distributed among hexamers. However, as $T$-number increases, there is a tendency for the stress to become lower at the quasi-sixfold axis. For $(n,0)$ capsids (i.e., $T=9,16,25$), the stress pattern is similar, and the stress is lower at the center of the faces of the icosahedron but higher along the edges connecting the pentamers. Similar, but not identical, energy distribution patterns have been found by Bowick *et al.* [28] and Altschuler *et al.* [29] in their study of Thomson problem, where electrons are interacting with a repulsive Coulomb potential on the surface of a sphere.

(4) Another issue, which is markedly different from the predictions of continuum elasticity theory, is the presence of local *shear and elongational* stresses on the hexamers in many of the $T$-number equilibrium structures. The *global* shear stress is zero, but some capsomers are under *local* shear stresses. This occurs whenever a capsomer has a different environment in different directions, as happens for those that do not lie at symmetry (rotation) axes. (In fact, the theory of "balanced structures" [30] shows that sites that lie on threefold or higher rotation axes are guaranteed to have zero tangential forces, by symmetry.) For instance, some hexamers in the $T=9$ capsid are in contact with one pentamer while the other hexamers do not have any pentamers as their neighbors. These differences of local environment give rise to the existence of local shear stresses. Nevertheless, shear stresses neutralize globally; otherwise, the capsid would be deformed and icosahedral symmetry would be destroyed.

(5) Another seemingly unexpected result is the presence of a "residual stress" in the optimal structures (i.e., the value of the lateral stress is not zero for the optimal radius). Figure 4 plots the values of average residual lateral stress—see Eq. (3)—for some small $T$-number optimal $(R=R^*)$ structures. The residual stress is positive, indicating an overall compression, and goes to zero as the radius of the capsid increases.

Naively, one would *a priori* expect that a fully equilibrated structure is characterized by a minimal energy *and zero stress*. According to the standard Laplace relation, in the absence of an imposed pressure, the lateral stress is expected to be zero. However, at the microscopic level this requirement is hard to reconcile with the geometry and nature of the system. In particular, in our system the discrete character and the special geometry of the capsids force the existence of residual lateral stress, which by no means implies that the capsid is mechanically unstable. In fact, in our system the residual lateral stresses are compensated by the radial components of the stress, in such a way that the global pressure difference is identically zero [see Eq. (12)]. Notice that, by definition, the pressure difference is zero at $R^*$, since $R^*$ is the optimal radius that minimizes the energy—and thus makes zero the right hand side of Eq. (7). As a consistency check, we have verified that the identity in Eq. (12) holds for all our optimal (energy minima) structures. It is worth mentioning that residual stresses have also been observed in another (covalently bound) nanoshell: carbon nanotubes [15].

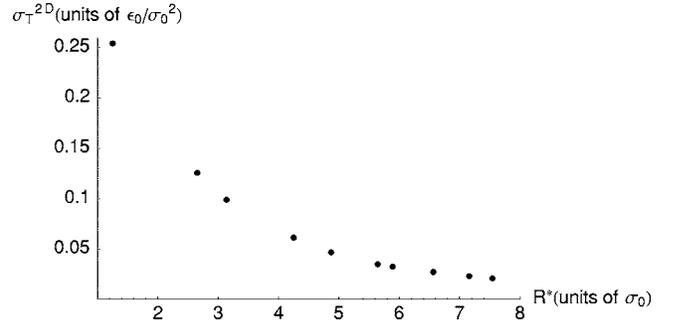

FIG. 4. 2D residual tangential stress as a function of the optimal radius for the smallest $T$-number structures.

### B. Radial stress

The distribution of the radial components of the stress is very similar to the lateral stress, since both are connected through Eq. (12). In particular, the following.

(1) The distribution of radial stress is also highly inhomogeneous, and strongly conditioned by the geometry of the capsid.

(2) The radial stress on pentamers is consistently higher than that on hexamers. The radial stress on pentamers of the equilibrium structures is typically positive (pointing outwards), while for the hexamers is negative (pointing inwards).

(3) The equilibrium structures have also a residual radial stress (i.e., the radial stress is not zero for the radius that optimizes the lateral energy), which is coupled to the residual lateral stress by Eq. (12). The radial stress of the optimal structures is negative and its value gets closer to zero as the capsid gets bigger. This is closely connected to the fact that, as the radius of the capsid increases, its surface gets flatter (less curvature), and thus the radial component of the force decreases.

Note that the mechanical behavior of these nanoshells is markedly different from the expectations of continuum elasticity theory of thin shells. In a spherical thin-walled vessel, for reasons of symmetry, the lateral stress must be uniform within the shell. Furthermore, shear stresses vanish identically at each point within the shell.

## V. MECHANICAL PROPERTIES OF STRAINED CAPSIDS

There are many influences that may lead to the swelling or compression of a capsid. Changes in $p$H, temperature or salt concentration, the application of external pressures, or the confinement of genetic or nongenetic materials inside a capsid could result in a change in its radius. Virus shrinkage has been observed in Semliki Forest Virus—a $T=4$ structure—upon exposure to low $p$H [31]. In recent years, virion swelling has been extensively studied for Satellite Mosaic Virus ($T=1$) and for many $T=3$ icosahedral plant viruses [32]. In particular, it has been shown by a variety of techniques—including high-resolution cryoEM and x-ray studies [13] and assays of the release and entry of viral RNA and other anionic polymers through capsids [9]—that changes in $p$H and $Ca^{2+}$ concentration can lead to significant swelling of CCMV. It is significant that this swelling—increased by more than 10% in the diameter of the capsid—involves only a radial expansion of the capsomers, with essentially no conformational change in the protein subunits or their relative positions within each pentamer or hexamer.

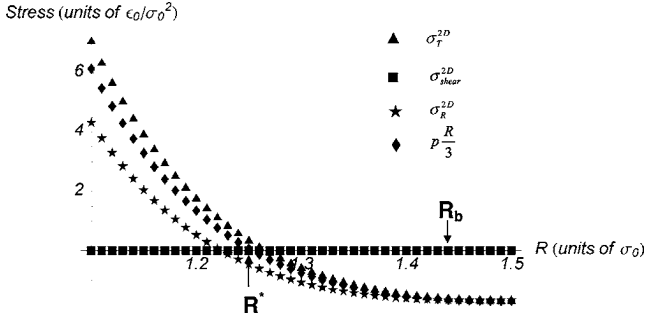

FIG. 5. 2D tangential (triangles), radial (stars), and lateral shear (squares) stress for a $T=1$ structure. Optimal and bursting radii are $R^*=1.25$ and $R_b=1.44$, respectively, (in units of $\sigma_0$).

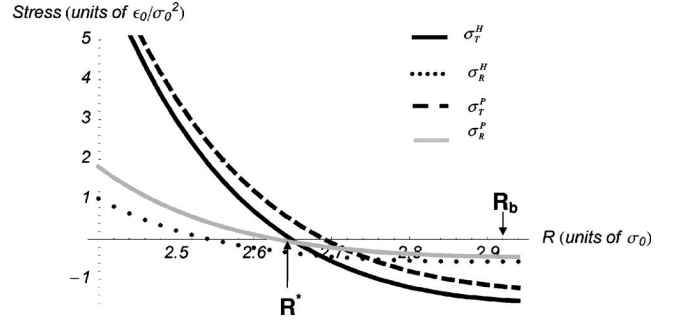

FIG. 6. Radial [see Eq. (17)] and 2D tangential [Eq. (15)] stresses as a function of capsid radius $R$, evaluated at hexamers (heavy and dotted lines, respectively) and pentamers (dashed and gray lines). Optimal and bursting radius are $R^*=2.65$ and $R_b=2.92$, respectively, (in units of $\sigma_0$).

The goal of this section is to examine the variations in the stress in response to changes in the radius of capsids. We assume that the strength of the interactions remains the same (although the aforementioned factors can also affect in general the value of $\varepsilon_0$). In particular, our aim is to investigate how the lateral and radial stresses in a capsid change in concert with each other, and what the maximum lateral stresses and pressure differences are that a capsid can withstand before breaking.

As noted in the previous section, the geometry of a capsid imposes an inhomogeneity in the stress distribution among capsomers, which as we will see renders many of the traditional continuum elasticity theory arguments inapplicable here.

### A. Homogeneous structure: $T=1$

For the sake of simplicity, we start with $N=12$ (i.e., $T=1$) structure. Figure 5 represents the local lateral, radial, and (45°) shear values of the stress in a $T=1$ structure, as a function of the radius of the capsid $R$ (in units of $\sigma_0$).

As shown in Fig. 5, the lateral stress at the optimal radius is slightly positive while the radial stress is negative. There is no shear stress at any point in the system. When $R$ increases beyond $R^*$, the capsid becomes swollen. At this point, the compressive lateral stresses become relieved, and lateral stretching appears throughout the capsid. Shear stress, by contrast, does not appear at any radius.

It should be emphasized that the $T=1$ structure has both residual lateral and radial stresses even at the optimal radius $R^*$. However, if we combine both the lateral and radial term in Eq. (12), we obtain that the pressure is identically zero at optimal $R^*$. This can be seen in Fig. 5 where the bold curve crosses zero. This is consistent with the fact that Monte Carlo simulations and the energy minimization were performed under zero imposed pressure.

Equation (12) enables us to calculate the pressure required to equilibrate the capsid under imposed strains (i.e., for $R$ different from $R^*$). For small $R$'s, the 2D stresses are positive indicating both lateral and radial compression. Through Eq. (12), this translates into a positive pressure difference on the capsid, i.e., an externally imposed pressure. Similarly, for $R>R^*$, the stress becomes negative indicating stretching, a situation that would occur under the influence of an internal pressure. Equation (12) also allows us to determine the distribution of stresses arising from a given imposed pressure. The values of the pressure corresponding to different lateral stresses are also depicted (see black curve) in Fig. 5 (properly scaled to have the same units as the 2D stress components).

### B. Inhomogeneous structures: $T>1$

For the other $T$-number structures, the situation becomes more complicated due to the presence of different types of capsomers. Figure 6 illustrates the lateral and radial values of the stress in a $T=3$ structure.

Both tangential and radial stresses are higher at pentamers than at hexamers. At the optimal radius for energy minimization $R^*$, hexamers have almost zero lateral stress and a negative radial stress, whereas pentamers are laterally compressed and slightly stretched in the radial direction.

For structures with two different types of capsomers, one could compute a "local" pressure for either pentamers or hexamers, by applying Eq. (12) locally, i.e.,

$$p^{P,H} = \frac{1}{R}(2\sigma_T^{P,H(2D)} + \sigma_R^{P,H(2D)}). \qquad (18)$$

Figure 7 plots the "local" pressure on both pentamers and hexamers vs the capsid radius $R$. We notice that at the opti-mal radius $R^*$, pentamers and hexamers are under different local pressures. Nevertheless, when we compute the actual pressure (including all capsomers), we find, again, that the total pressure vanishes at the optimal radius. Note that, in our model, only the global pressure has meaning. We, indeed, force the capsomers to live on the surface of a "rigid" sphere, and we do not allow pentamers and hexamers to move in the radial direction independent from each other. Note that both local pressure and the radial components of the stress are higher in pentamers than in hexamers, according to Figs. 5 and 6. This suggests that in a nanoindentation experiment, using an AFM tip with dimensions much smaller than the radius of the capsid (such as a nanotube tip), one would expect to find that pentamers react differently from hexamers. At present, experiments [33,8,34] are not quite able to probe these effects in detail.

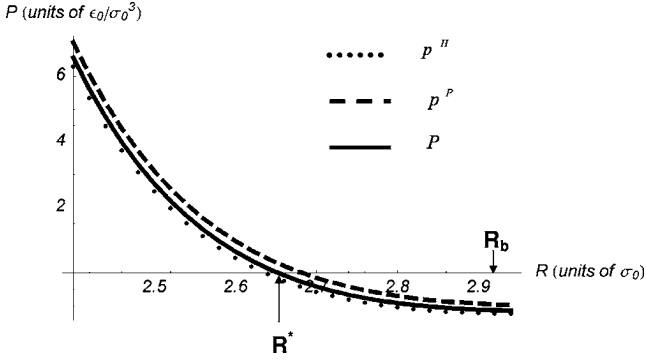

FIG. 7. Local values of the pressure in the hexamers (dotted line), the pentamers (dashed line), and the global pressure (black line) as a function of capsid radius $R$. Optimal and bursting radius are $R^*=2.65$ and $R_b=2.92$, respectively.

For higher $T$-number structures, capsomers have different stress levels, and the inhomogeneity of the stress complicates the analysis even further. However, the trends in behavior are qualitatively similar to that of $T=3$.

### C. Elastic moduli of capsids

From the plots of pressure vs radius $R$, one can extract one of the most important quantities characterizing the mechanical properties of nanoshells: the bulk modulus. This quantity can also be obtained directly from the energy vs radius plots resulting from our simulations. The bulk modulus $K$ gives the change in pressure of a substance as its volume changes. It is defined as

$$K = -V \left(\frac{\partial P}{\partial V}\right)_T. \qquad (19)$$

Considering a capsid as a spherical object of volume $V=4\pi R^3/3$, its effective bulk modulus, according to Eq. (19), would be given by

$$K = -\frac{R}{3} \left(\frac{\partial P}{\partial R}\right)_T. \qquad (20)$$

Figure 8 shows the plot of the effective bulk modulus, Eq. (20), as a function of the radius for a $T=3$ capsid. We ob-serve that as the radius of the capsid increases, the bulk modulus decreases implying that swollen capsids are "softer" than optimal or compressed capsids.

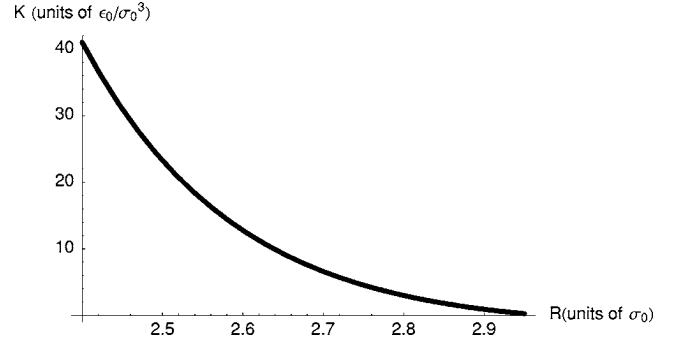

FIG. 8. Bulk modulus of a $T=3$ capsid as a function of capsid radius $R$.

We can also make a numerical estimate of the bulk modulus for an optimal $T=3$ capsid (i.e., $R=R^*$). From Fig. 6, we obtain that the value of bulk modulus at $R^*$ is $K=9.1\ \varepsilon_0/\sigma_0^3$. Considering a capsid with radius of $R=26$ nm (roughly that of CCMV), and using the estimate of the strength of the capsomer-capsomer interaction of $15kT$, the value of the bulk modulus would be roughly $5\times 10^6$ Pa (i.e., 5 MPa).

### VI. BURSTING AND MECHANICAL FAILURE OF THE CAPSID

Both expansion and compression of a capsid may eventually lead to its mechanical failure. In the previous section, we observed that as $R$ is increased beyond $R^*$, tangential stress develops throughout the capsid. Eventually, this stress may become too large for a capsid to tolerate and a crack appears. A set of very interesting questions then arises: When will the capsid fail? What are the conditions that control the mechanical failure of capsids? What is the maximal internal pressure that a capsid can resist?

One can use a very simple (but approximate) argument to get some insight into the limits of failure of capsids. In our model, capsomers are held together by an intermolecular Lennard-Jones potential [Eq. (1)]. The maximum force that two bound capsomers can resist before their complete separation is that at the inflection point of the potential:

$$r_{\text{inf}} = \sigma\left(\frac{13}{7}\right)^{1/6}. \qquad (21)$$

Thus the swelling of a capsid is expected to lead to its failure via the formation of a crack whenever the distance between capsomers exceeds the inflection point of the potential, $r_{\text{inf}}$. This would occur roughly at a capsid radius

$$R_b \approx R^*\left(\frac{13}{7}\right)^{1/6}, \qquad (22)$$

which corresponds approximately to a 10% expansion in radius.

The above argument indicates that the maximum expansion that a capsid may tolerate depends mainly on the range of attractive interactions between its capsomes rather than on their strength of attraction. Short-range interactions between rigid capsomers will not allow a significant expansion of a capsid before bursting. Thus, the significant ability of capsids for swelling may reveal the presence of relatively long-range effective interactions among the capsomers. In the case of CCMV [7], the flexibility of the C terminals of the capsid proteins, which mediate the intercapsomer interaction, can be the responsible for the long-range nature of the effective interactions [35].

It is also possible to make a very crude estimate of the maximum lateral stress that a capsid can tolerate from the force imposed on the system and the distance between capsomers at the inflexion point. For a flat 2D hexagonal network of Lennard-Jones particles with nearest-neighbor interactions and subjected to a homogeneous strain, the stress at the inflection point of the potential turns out to be

$$\sigma_T^{2D}|_b \approx -5\frac{\varepsilon_0}{\sigma^2}. \quad (23)$$

This sets a very crude estimate of the order of magnitude of the maximum lateral stress tolerable for our model capsid. Of course this estimate is only approximate and is expected to become worse as the curvature of the capsid increases. It also ignores the presence of fivefold defects.

All the previous arguments are valid only at zero temperature. At finite temperature, the energy of thermal fluctuations will lead to bursting of the capsid before the inflection point and/or before the critical $\sigma_T^{2D}$ value estimated in Eq. (23) are reached.

In order to get more insight into the effect of strain on capsids structures, we have performed Monte Carlo simulations to study the bursting of different $T$-number capsids. For a given number of capsomers, $N$, we first equilibrate our system at $R=R^*$. Using this equilibrated structure as a starting configuration of a Monte Carlo simulation, we let the structure evolve at reduced temperature of $T=0.01$ for a fixed number of MC steps (corresponding to a fixed amount of time). We repeat the simulation for increasing values of the radius $R$ and finally find for which radius $R_b$ the structure is first disrupted. Note that the radius $R_b$—at which bursting occurs—depends sensitively on temperature. The appearance of a crack is similar to a first order transition: when $R$ exceeds a critical value $R_b$, first a tiny crack appears but then the capsid bursts dramatically with a large crack stretching across the capsid surface. Figure 9 illustrates some cracked capsids with different $T$ numbers.

Figure 10 shows the plot of the radius at which bursting occurs vs the optimal radius $R^*$ for different $T$ structures. We observe a linear dependence with a slope of 1.05, which indicates that these nanoshells can tolerate a 5% radial expansion (that is a 15% expansion in volume) before bursting. Significant swelling, which in some cases results in a 10% radial expansion, has been observed experimentally for some viruses, such as CMMV [7], for which the flexibility of C terminals has been suggested to play an important role [35].

Except for structures with very small $T$ number, we observe that capsid brusting occurs when the maximum value of the local lateral stress reaches a value close $-5\varepsilon_0/\sigma^2$ which is, in turn, surprisingly close to the approximate zero-temperature prediction of Eq. (23). It is quite interesting that a very similar value of $-4.5\varepsilon_0/\sigma^2$ also appears in previous studies of the crack formations in (flat) 2D Lennard-Jones particles [36,37]. In these studies, it was found that at low temperatures ($\varepsilon \ll k_BT$) a 2D crystal—free of any defects—failed when the applied stress reached a value of $-4.5(\varepsilon/\sigma^2)$. Just before a crack opened up the 2D stress was calculated to be $-4.5$ at all sites throughout the lattice. In the presence of vacancies or defects, the system failed at lower values of the *global* 2D stress. However, the maximum *local* stress, just around the vacancy or defect, reached also a value of $-4.5(\varepsilon/\sigma^2)$ at the onset of failure. Consequently, the value of $-4.5(\varepsilon/\sigma^2)$ seems to be an *intrinsic* local property of 2D Lennard-Jones systems, with and without defects. A careful examination of the bursting of our capsids suggests that a similar conclusion holds as well for (big enough) closed icosahedral shells and confirms that this kind of failure is a generic feature of localized defects and short-ranged interactions.

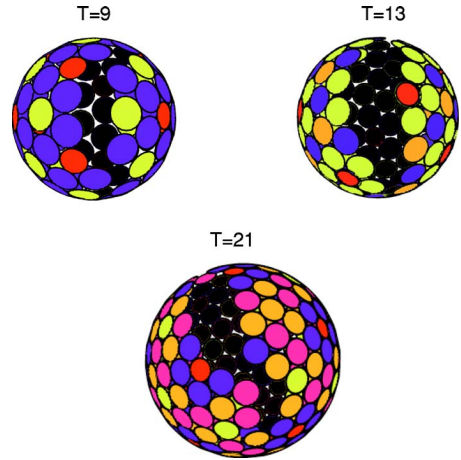

FIG. 9. (Color online) Cracks developed in different $T$-number capsids.

Another important issue is to determine the weakest point of the structure. Figure 3 illustrates that the largest negative stress (stretching) is present at hexamers. Therefore, under expansion, capsid failure will occur at the contact between hexamers. The snapshots of the simulations of capsid bursting—see Fig. 9—are consistent with this observation.

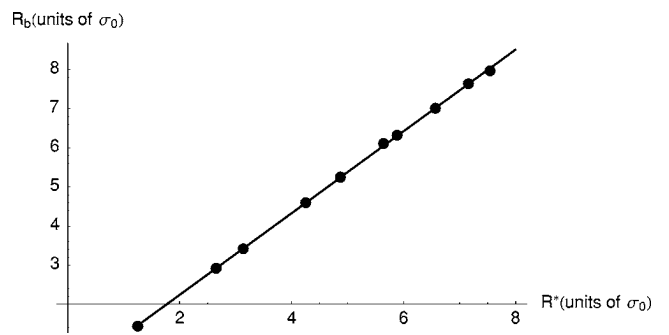

FIG. 10. Radius at capsid burst vs their optimal radius in units of $\sigma_0$. The line is a fit with slope 1.05.

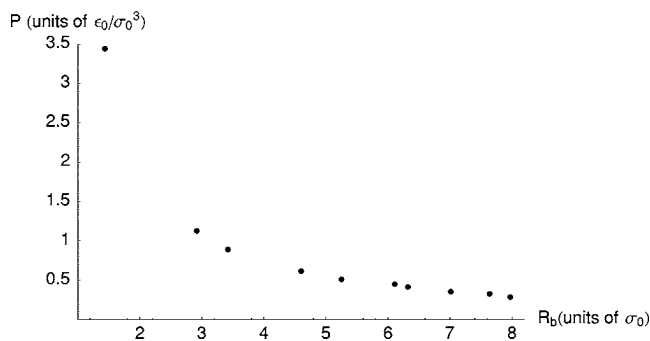

FIG. 11. Maximum internal pressure that different $T$-number capsids can tolerate vs the bursting radius.

In general, the study of capsid bursting helps us learn about the maximum internal pressure or the maximum load that these nanoshells can tolerate before they crack.

Figure 11 plots pressure vs bursting radius for some of the $T$-number structures. As expected, as the capsids get larger, the maximum pressure that they can tolerate becomes smaller and eventually goes to zero.

## VII. CONCLUSIONS

In this paper we have investigated the unique mechanical properties of viral capsids. We have shown that the symmetry of the capsid is a very important factor in determining its mechanical properties: the distribution of stress is in general highly inhomogeneous, and continuum elasticity arguments are not necessarily applicable in the description of one-molecule-thick, nanometer-sized, viral shells. Furthermore, we have characterized the radial and lateral components of the stress in viral capsids, and have shown that these stresses are directly connected to the distribution of disclinations and to internal and external pressures on capsids [see Eq. (12)].

One important underlying reason for the study of the mechanical properties of viruses is to improve our understanding of the disassembly processes associated with the different scenarios for release of the genetic material. In Sec. VI, we analyzed the bursting or "in-plane" failure of capsids. This is a *mechanical* failure occurring when the attractive forces holding capsomers together are no longer strong enough to resist the lateral forces imposed by, for instance, an internal pressure. In that situation, the capsid relieves its in-plane stress via crack formation initiated at a hexameric site. This bursting of the capsid is, indeed, one plausible scenario for the release of genetic material.

A second mechanism of failure—"thermodynamic" disassembly—will take place, however, when the capsid becomes energetically unfavorable compared to having the proteins in solution. The imposition of an internal or external pressure or the presence of a bending energy introduces additional terms into the energetic balance of a capsid and may eventually render a capsid thermodynamically unstable. In general, the threshold for disassembly can be determined by calculating the radius at which the energy $E_i$ per capsomer—including the effects of pressure and bending—exceeds the chemical potential of capsomers in solution.

In the situation where the capsid becomes thermodynamically unfavorable, pentamers would be the first to leave, since they have the highest energy (they only have 5 contacts vs 6 of the hexamers). Furthermore, we have seen in the previous section that pentamers have typically the highest compressive stress. Therefore it is plausible that the disassembly would be initiated by pentamer release. This scenario has, in fact, been reported experimentally in the case of the TYMO (turnip yellow mosaic) virus and P22, where a change in ambient conditions leads to popping out of one (or all) of the 12 fivefold (pentameric) capsomers [38,39]. This "decapsidation" leads also to the release of the genetic material inside the capsid, and it is thus a likely scenario for gene delivery.

Thermodynamic failure will prevail over mechanical instability if the above threshold is reached before the critical bursting radius $R_b$. The fact that the chemical potential can be controlled by solution conditions ($p$H, salt, etc.) and protein concentration opens up the possibility of favoring the disassembly vs the lateral bursting of a capsid in an experiment.

It will be interesting to see if our predictions about the bursting of a capsid can be probed in osmotic shock experiments. These experiments should allow one to measure the maximum pressure that a capsid can sustain before it fails. For instance, if a capsid is self-assembled from its proteins constituents in the presence of PEG—polyethylene glycol—(or other osmolytes), and then placed in buffer solution with *no* PEG, it will be possible to measure the extent to which a capsid can be pressurized. The results of these experiments would allow us to infer the strength and the range of the capsomer-capsomer potential. The advantage of the capsid expansion by osmolytes over a $p$H expansion is that changes in the $p$H can also modify the interaction potential between protein subunits, and thereby complicate the interpretation of the results.

Finally, evaluation of the maximum strains and pressures sustained by viruses provide important estimates of their capacity to encapsidate different lengths of genetic material. This has important implications for non-genetic biomedical applications as well as gene therapy. Mechanical properties of these nanohedral shells are clearly important for understanding both the biophysical function of viral capsids, and the technological potential of these biomaterials as efficient nanocontainers.

## ACKNOWLEDGMENTS

We have benefited from many helpful discussions with Professor Robijn Bruinsma, Professor William M. Gelbart, and Professor Howard Reiss. This work was supported by Grant No. NSFCHE-03013563 and No. NSFCHE04-00363. D.R. acknowledges support by the Ministerio de Ciencia y Tecnología of Spain through the "Ramón y Cajal" program.

## APPENDIX: PRESSURE AND 3D VIRIAL STRESS TENSOR

In this appendix, we derive the relation between the 3D virial stress tensor and the thermodynamic pressure. We fol-low the dimensional argument of Bogoliubov and Green [40] for our special situation.

We start from the partition function of a spherical capsid

$$Q(N,V,T) = \frac{1}{\Lambda^{3N} N!} \int d\mathbf{r}_1 \int d\mathbf{r}_2 \int d\mathbf{r}_N e^{-\beta U(\{r_i\})}, \quad (A1)$$

where integration is carried out over all space, $\beta = 1/kT$, $\Lambda$ is the de Broglie wavelength, and the potential energy $U(\{r_i\}) = \frac{1}{2}\Sigma_{i,j} V(r_{ij})$ represents the sum of all capsomer interactions.

The pressure is defined thermodynamically as

$$p = -\left(\frac{\partial F}{\partial V}\right)_{N,T} = kT\left(\frac{\partial \ln Q}{\partial V}\right)_{N,T}. \quad (A2)$$

For a spherical object, such as our capsids, this means

$$p = -\left(\frac{\partial F}{\partial V}\right)_{N,T} = -\frac{1}{4\pi R^2}\left(\frac{\partial F}{\partial R}\right)_{N,T}. \quad (A3)$$

Let us rescale all the positions of capsomers by the radius $R$ of the capsid

$$\mathbf{r}_i = R\mathbf{t}_i, \quad (A4)$$

where the new scaled coordinates $\mathbf{t}_i$ now range from $-1$ to $1$. The partition function then becomes

$$Q(N,V,T) = \left(\frac{R}{\Lambda}\right)^{3N} \frac{1}{N!} \int d\mathbf{t}_1 \int d\mathbf{t}_2 \int d\mathbf{t}_N e^{-\beta U(\{R\mathbf{t}_i\})} \quad (A5)$$

and its derivative with respect to the volume yields

$$\frac{\partial \ln Q}{\partial V} = \frac{1}{Q}\frac{1}{4\pi R^2}\frac{\partial Q}{\partial R}$$

$$= \frac{N}{V} - \frac{1}{Q}\frac{1}{4\pi R^2}\left(\frac{R}{\Lambda}\right)^{3N}\frac{1}{N!}\int d\mathbf{t}_1 \int d\mathbf{t}_2 \int d\mathbf{t}_N$$

$$\times e^{-\beta U(\{\mathbf{t}_i\})}\left(\frac{1}{2}\sum_{i,j}\frac{\partial V(r_{ij})}{\partial r_{ij}}\frac{\partial r_{ij}}{\partial R}\right). \quad (A6)$$

With

$$\frac{\partial r_{ij}}{\partial R} = \frac{1}{R}\frac{(x_j - x_i)^2 + (y_j - y_i)^2 + (z_j - z_i)^2}{r_{ij}} \quad (A7)$$

and Eq. (A6) (returning now to the original coordinates), one gets

$$\frac{\partial \ln Q}{\partial V} = \frac{N}{V} - \frac{1}{3V}$$

$$\times \left\langle \frac{1}{2}\sum_{i,j}\frac{\partial V(r_{ij})}{\partial r_{ij}}\frac{(x_j - x_i)^2 + (y_j - y_i)^2 + (z_j - z_i)^2}{r_{ij}}\right\rangle, \quad (A8)$$

where the braquets $\langle \rangle$ denote equilibrium ensemble averages. Defining the 3D virial stress tensor as

$$\sigma^{3D}_{\alpha\beta} = \frac{NkT}{V} - \frac{1}{2}\sum_{i,j}\frac{dV(r_{ij})/dr_{ij}}{r_{ij}}(r_{ij}\cdot\hat{e}_\alpha)(r_{ij}\cdot\hat{e}_\beta), \quad (A9)$$

where $\alpha,\beta$ denote the indices in Cartesian coordinates $(x,y,z)$, and $\hat{e}_\alpha, \hat{e}_\beta$ are the unit vectors along the $x,y,z$ directions, Eq. (A8) can be rewritten as

$$p = \left\langle \frac{1}{3}\text{Tr }\sigma^{3D}\right\rangle, \quad (A10)$$

where Tr stands for the trace of the stress tensor. Since the trace of a tensor is invariant with respect to coordinate changes, we can express the previous relation in spherical coordinates as

$$p = \left\langle \frac{1}{3}\text{Tr}\sigma^{3D}\right\rangle$$

$$= \left\langle \frac{1}{3}(\sigma^{3D}_{\theta\theta} + \sigma^{3D}_{\varphi\varphi} + \sigma^{3D}_{rr})\right\rangle$$

$$= \left\langle \frac{2\sigma^{2D}_T}{R} + \frac{1}{3}\sigma^{3D}_{rr}\right\rangle, \quad (A11)$$

which is Eq. (8) in the text, and where, in the last step, the expression for the tangential 2D stress has been introduced.